\documentclass[conference, 10pt]{IEEEtran}	

\IEEEoverridecommandlockouts

\usepackage{algorithm,algorithmic,amsmath,amssymb,amsthm,array,bibentry,cite,color,comment,enumerate,enumitem,eurosym,float,graphicx}
\usepackage{lettrine,mathrsfs,multicol,multirow,nomencl,pict2e,psfrag,ragged2e,setspace,stfloats,subfigure,supertabular,tabularx,url}
\usepackage[USenglish]{babel}
\usepackage[absolute]{textpos}

\setlist[itemize]{leftmargin=5mm}

{		\newtheorem{theorem}{Theorem}}
{ 		}
{ 		}
{ 		}
{ 		\newtheorem{remark}{Remark}}
{		}
{		}
{ 		\newtheorem{corollary}{Corollary}}
{		

\renewcommand{\b}{\mathbf{b}}

\newcommand{\g}{\mathbf{g}}
\newcommand{\h}{\mathbf{h}}

\renewcommand{\u}{\mathbf{u}}
\renewcommand{\v}{\mathbf{v}}

\newcommand{\x}{\mathbf{x}}

\newcommand{\z}{\mathbf{z}}

\newcommand{\0}{\mathbf{0}}

\newcommand{\I}{\mathbf{I}}

\newcommand{\N}{\mathbf{N}}

\newcommand{\Y}{\mathbf{Y}}

\newcommand{\setC}{\mathcal{C}}

\newcommand{\setE}{\mathcal{E}}

\newcommand{\setI}{\mathcal{I}}

\newcommand{\setL}{\mathcal{L}}

\newcommand{\setN}{\mathcal{N}}

\newcommand{\setV}{\mathcal{V}}

\newcommand{\Real}{\mbox{$\mathbb{R}$}}
\newcommand{\Compl}{\mbox{$\mathbb{C}$}}

\newcommand{\Exp}{\mathbb{E}}
\newcommand{\herm}{\mathrm{H}}

\begin{document}

\begin{textblock}{14}(1,.5)
\begin{center}
Paper presented at the IEEE ICC 2015 - Workshop on 5G \& Beyond - Enabling Technologies and Applications \\ ('ICC'15 - Workshops 23').
\end{center}
\end{textblock}

\title{Fractional Pilot Reuse in Massive MIMO Systems \\ \vspace{4mm}}

		
\author{\IEEEauthorblockN{Italo Atzeni, Jes\'{u}s Arnau, and M\'{e}rouane Debbah}%
		\IEEEauthorblockA{Mathematical and Algorithmic Sciences Lab, France Research Center, Huawei Technologies Co. Ltd. \\
						  \{italo.atzeni, jesus.arnau, merouane.debbah\}@huawei.com}}

\maketitle

\begin{abstract}
Pilot contamination is known to be one of the main impairments for massive MIMO multi-cell communications. Inspired by the concept of fractional frequency reuse and by recent contributions on pilot reutilization among non-adjacent cells, we propose a new pilot allocation scheme to mitigate this effect. The key idea is to allow users in neighboring cells that are closest to their base stations to reuse the same pilot sequences. Focusing on the uplink, we obtain expressions for the overall spectral efficiency per cell for different linear combining techniques at the base station and use them to obtain both the optimal pilot reuse parameters and the optimal number of scheduled users. Numerical results show a remarkable improvement in terms of spectral efficiency with respect to the existing techniques.
\end{abstract}

\begin{IEEEkeywords}
Channel estimation, fractional frequency reuse, fractional pilot reuse, massive MIMO, pilot contamination.
\end{IEEEkeywords}

\section{Introduction} \label{sec:Intro}

Massive multiple-input multiple-output (MIMO) refers to a type of cellular network based on multiuser MIMO in which the number of antennas at the base station (BS) is much larger than the number of served user terminals \cite{Mar10,Rus13,Boc13}. Under certain propagation conditions, such a system renders quasi-orthogonal channels among users and very simple linear processing is shown to be optimal in terms of throughput.

In this context, a problem that appears when obtaining channel state information is that the number of available pilot sequences is finite and limited by the channel behavior, as their duration cannot span larger than the coherence interval of the channel to be estimated. In consequence, mobile users in different cells might have to reuse the same pilot sequences, thus resulting into corrupted channel estimates at each BS: this is known as the \emph{pilot contamination} effect \cite{Jos11} and represents a major impairment affecting massive MIMO communications.

This problem has been intensively investigated in the literature in the last few years. For instance, \cite{Yin13} proposed a low-rate coordination phase between different BSs; \cite{Zha14} introduced an elaborated scheme for OFDM-based massive MIMO systems, with differentiated downlink and uplink scheduled training phases; \cite{Mul14} proposed a subspace projection to improve the channel estimation accuracy; and \cite{New14} also exploited coordination, comparing the amount of training needed with respect to the uncoordinated case and its implications.

More recently, the idea of pilot reuse was introduced in \cite{Huh12}. In this respect, \cite{Bjo14,Bjo15} advocated for employing orthogonal pilot subsets in adjacent cells, optimizing the number of required subsets and the number of scheduled users per cell that maximizes the overall spectral efficiency, for both uplink and downlink. This idea is borrowed from the well-established concept of frequency reuse in cellular systems \cite{Don79} and is therein coined as ``fractional pilot reuse'': given an integer $\beta \geq 1$ representing the number of orthogonal pilot subsets (or, equivalently, the ratio between the number of available sequences and the number of scheduled users per cell), for each cell, only a fraction $1/\beta$ of the interfering cells reuse the same pilot subset.

\begin{figure}[t!]
\centering
\includegraphics[scale=1]{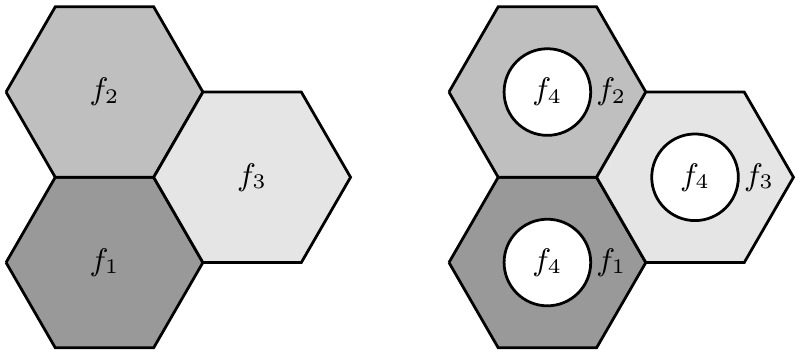}
\caption{Traditional frequency reuse with $\alpha=3$ sub-bands and the equivalent FFR setup.} \label{fig:FFR}
\end{figure}

On the other hand, the concept of fractional frequency reuse (FFR) consists in splitting the available bandwidth into $\alpha+1$ frequency sub-bands, where the integer $\alpha \geq 1$ specifies the frequency reuse factor among the cell edges and the additional sub-band is reused in all the cell centers (see \cite{Nov11} and references therein). The difference between traditional frequency reuse and FFR is illustrated in Figure~\ref{fig:FFR}.


The idea behind this paper is to adapt the aforementioned concept of FFR to the pilot domain and assign the pilot sequences to the users within the cells in the same way as different frequency sub-bands are allocated in FFR. Under these premises, we propose a novel channel estimation scheme where a fraction of users within each cell reuses the same pilot subset across the whole system, while the rest are allocated orthogonal subsets depending on a reuse parameter. We also refer to this approach as \emph{fractional pilot reuse}, but in the sense that a \emph{fraction} of the users per cell reuse the same pilot subset in all the cells. In this regard, we derive closed-form expressions for the spectral efficiency and analyze the performance in the large-antenna regime. Numerical results show that such deployment outperforms the proposal in \cite{Bjo14} and offers remarkably good throughput, especially for large numbers of antennas at the BS.

The rest of the paper is structured as follows. In Section~\ref{sec:system_model}, we introduce the system model on which we build our novel framework based on fractional pilot reuse in Section~\ref{sec:fractional_pilot_reuse}. In Section~\ref{sec:numerical_results}, numerical results are reported to corroborate the proposed scheme. Finally, Section~\ref{sec:conclusions} draws some concluding remarks.

\section{System Model} \label{sec:system_model}

Let us consider a massive MIMO cellular network over a set of cells denoted by $\setL$. The BS of each cell $l \in \setL$ is equipped with $N$ antennas and serves $K$ single-antenna users in the uplink; in this context, $N$ is fixed while $K$ is chosen adaptively. We model the channels as block-fading, where $\h_{j}(\z_{lk}) \in \Compl^{N}$ represents the channel response between BS $j$ and user $k$ in cell $l$, with random coordinates $\z_{lk} \in \Real^{2}$, during its coherence block consisting of $T_{lk}$ channel uses: in the following, we consider that the coherence block of the whole system spans $T \triangleq \min_{l,k} \{ T_{lk} \}$ channel uses.\footnote{A user with high mobility, which would consistently decrease the overall coherence block $T$, can be served in a different time-slot in order not to compromise the channel estimation of the other users.} The channel realizations are drawn from a circularly symmetric complex Gaussian distribution, i.e., $\h_{j}(\z_{lk}) \sim \setC \setN \big( \0, d_{j}(\z_{lk}) \I_{N} \big)$, where $d_{j}(\z_{lk})$ represents the variance of the channel attenuation from BS $j$ to user $k$ in cell $l$ (cf. \cite{Bjo15}): this term accounts for propagation impairments such as pathloss and shadowing.

Let us introduce $\Compl^{T} \ni \x_{lk} \triangleq \big( x_{lk}(t) \big)_{t=1}^{T} $ as the vector signal transmitted by user $k$ in cell $l$, normalized as $\Exp \big\{ |x_{lk}(t)|^{2} \big\} = 1$, $\forall t \in [1,T]$. The received signal at BS $j$ in a coherence block, denoted by $\Y_{j} \in \Compl^{N \times T}$, is thus given by
\begin{equation} \label{eq:Y}
\Y_{j} \triangleq \sum_{l \in \setL} \sum_{k=1}^{K} \sqrt{p_{lk}} \h_{j}(\z_{lk}) \x_{lk}^{\herm} + \N_{j}
\end{equation}
where $p_{lk}$ is the transmit power allocated to user $k$ in cell $l$ and $\N_{j} \in \Compl^{N \times T}$ is the additive noise at BS $j$, whose elements are distributed independently as $\setC \setN (0, \sigma^{2})$. As in \cite{Bjo15}, we adopt a statistic-aware power control,\footnote{Such power-control policy makes the average effective channel gain equal for all users.} where the signals from user $k$ in cell $l$ are allocated the power $p_{lk} = \rho/d_{l}(\z_{lk})$, with $\rho > 0$.

We take into account a pilot book $\setV$ of $B \in [1, T]$ orthogonal pilot sequences defined as
\begin{equation}
\setV \triangleq \{ \v_{i} \}_{i=1}^{B}, \qquad \; \v_{a}^{\herm} \v_{b} = \left\{
\begin{array}{ll}
\hspace{-1mm} B, & \textrm{if} \ a=b \\
\hspace{-1mm} 0, & \textrm{otherwise}
\end{array} \right.
\end{equation}
with each $\v_{i} \in \Compl^{B}$; in the following, we will use $i_{lk}$ to denote the index of the pilot sequence assigned to user $k$ in cell $l$. Therefore, the vector signal of user $k$ in cell $l$ can be written as $\x_{lk} \triangleq (\u_{lk}, \v_{i_{lk}})$, where $\u_{lk} \in \Compl^{T-B}$ represents the information vector. We can thus decompose the received signal in \eqref{eq:Y} as $\Y_{j} = (\widehat{\Y}_{j} \; \widetilde{\Y}_{j})$ with
\begin{align}
\widehat{\Y}_{j} & \triangleq \sum_{l \in \setL} \sum_{k=1}^{K} \sqrt{p_{lk}} \h_{j}(\z_{lk}) \u_{lk}^{\herm} + \widehat{\N}_{j} \\
\widetilde{\Y}_{j} & \triangleq \sum_{l \in \setL} \sum_{k=1}^{K} \sqrt{p_{lk}} \h_{j}(\z_{lk}) \v_{lk}^{\herm} + \widetilde{\N}_{j}
\end{align}
with $\N_{j} = (\widehat{\N}_{j} \; \widetilde{\N}_{j})$. Moreover, we assume that BS $j$ applies a linear receive combining vector $\g_{jk} \in \Compl^{N}$ to the received signal, i.e., $\g_{jk}^{\herm} \widehat{\Y}_{j}$, so as to amplify the signal coming from its user $k$ while rejecting the interference caused by the others. In this respect, we consider:
\begin{itemize}
\item[a)] \emph{maximum ratio combining} (MRC), i.e., a traditional passive rejection scheme;
\item[b)] \emph{pilot-based zero-forcing combining} (P-ZFC), recently proposed in \cite{Bjo15}, which actively suppresses both intra- and inter-cell interference.
\end{itemize}

\section{Fractional Pilot Reuse} \label{sec:fractional_pilot_reuse}

\begin{figure}[t!]
\centering
\includegraphics[scale=1]{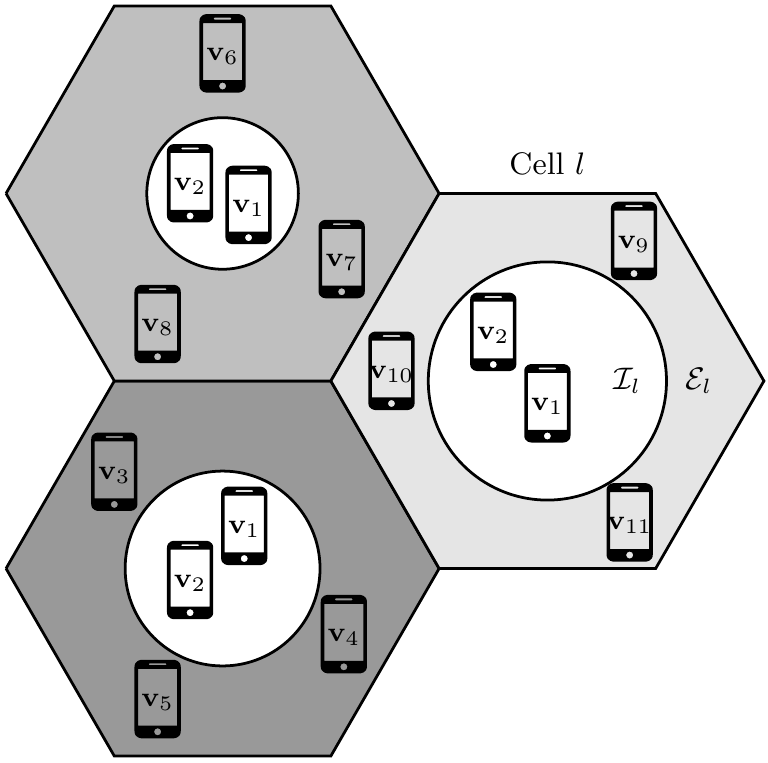}
\caption{Example of fractional pilot reuse (FPR) with $K=5$, $\beta=3$, and $\beta_{\mathrm{f}}=2/K$: in this setting, $B=11$ orthogonal pilot sequences are required.} \label{fig:cells}
\end{figure}

\subsection{Problem Formulation}\label{subsec:problem_formulation}


Inspired by the concept of FFR, we split the pilot book $\setV$ into $\beta+1$ subsets, where the integer $\beta \geq 1$ is referred to as \emph{integer reuse factor}. The users that, inside each cell $l$, exhibit a low variance of the channel attenuation with respect to their BS (i.e., $d_{l}(\cdot)$) reuse the $\beta$ subsets in non-adjacent cells; on the other hand, the users that, inside each cell $l$, enjoy a high $d_{l}(\cdot)$ are allocated the same $(\beta+1)$-th subset across the whole system.\footnote{In Section~\ref{sec:numerical_results}, we consider a standard pathloss model in which the users are divided into two groups simply based on the distance from their BS.} Hence, we define the new parameter $\beta_{\mathrm{f}} \in [0,1)$ as the \emph{fractional reuse factor}, which represents the fraction of users per cell that are assigned the additional pilot subset. We refer to this approach as \emph{fractional pilot reuse} (FPR): an illustrative example is shown in Figure~\ref{fig:cells}. The number of pilot sequences (and, hence, their length) is determined as
\begin{equation} \label{eq:FPR}
B \triangleq K \big( \beta_{\mathrm{f}} + (1 - \beta_{\mathrm{f}}) \beta \big).
\end{equation}
Note that $\beta_{\mathrm{f}} = 1$ would imply assigning the same $B=K$ sequences to each cell with no pilot reuse whatsoever (this case is prevented by the definition of $\beta_{\mathrm{f}}$). On the other hand, if $\beta_{\mathrm{f}} = 0$, the formulation trivially collapses to $B = \beta K$ as in the baseline \cite{Bjo15}.

\begin{figure*}[t!]
\addtocounter{equation}{+3}
\begin{align}
\label{eq:SINR_i_mrc} \mathsf{SINR}_{\setI_{j}}^{\mathrm{MRC}} & \triangleq \frac{B}{B \sum \limits_{l \in \setL \backslash \{j\}} \Big( \mu_{j, \setI_{l}}^{(2)} + \frac{\mu_{j, \setI_{l}}^{(2)} - (\mu_{j, \setI_{l}}^{(1)})^{2}}{N} \Big) + \Big( \sum \limits_{l \in \setL} \mu_{j, \setI_{l}}^{(1)} \frac{K}{N} + \frac{\sigma^{2}}{N \rho} \Big) \Big( B \sum \limits_{l \in \setL} \mu_{j, \setI_{l}}^{(1)} + \frac{\sigma^{2}}{\rho} \Big)} \\
\label{eq:SINR_e_mrc} \mathsf{SINR}_{\setE_{j}}^{\mathrm{MRC}} & \triangleq \frac{B}{B \sum \limits_{l \in \setL_{j} \backslash \{j\}} \Big( \mu_{j, \setE_{l}}^{(2)} + \frac{\mu_{j, \setE_{l}}^{(2)} - (\mu_{j, \setE_{l}}^{(1)})^{2}}{N} \Big) + \Big( \sum \limits_{l \in \setL_{l}} \mu_{j, \setE_{l}}^{(1)} \frac{K}{N} + \frac{\sigma^{2}}{N \rho} \Big) \Big( B \sum \limits_{l \in \setL_{j}} \mu_{j, \setE_{l}}^{(1)} + \frac{\sigma^{2}}{\rho} \Big)} \\
\label{eq:SINR_i_zfc} \mathsf{SINR}_{\setI_{j}}^{\mathrm{P-ZFC}} & \triangleq \frac{B}{B \sum \limits_{l \in \setL \backslash \{j\}} \Big( \mu_{j, \setI_{l}}^{(2)} + \frac{\mu_{j, \setI_{l}}^{(2)} - (\mu_{j, \setI_{l}}^{(1)})^{2}}{N-B} \Big) + \frac{\beta_{\mathrm{f}} K}{N-B} \bigg( \sum \limits_{l \in \setL} \mu_{j, \setI_{l}}^{(1)} \bigg( 1 - \frac{\mu_{j, \setI_{l}}^{(1)}}{\sum \limits_{\ell \in \setL} \mu_{j, \setI_{\ell}}^{(1)} + \frac{\sigma^{2}}{B \rho}} \bigg) \bigg) \Big( B \sum \limits_{l \in \setL} \mu_{j, \setI_{l}}^{(1)} + \frac{\sigma^{2}}{\rho} \Big)} \\
\label{eq:SINR_e_zfc} \mathsf{SINR}_{\setE_{j}}^{\mathrm{P-ZFC}} & \triangleq \frac{B}{B \sum \limits_{l \in \setL_{j} \backslash \{j\}} \Big( \mu_{j, \setE_{l}}^{(2)} + \frac{\mu_{j, \setE_{l}}^{(2)} - (\mu_{j, \setE_{l}}^{(1)})^{2}}{N-B} \Big) + \frac{(1 - \beta_{\mathrm{f}}) K}{N-B} \bigg( \sum \limits_{l \in \setL_j} \mu_{j, \setE_{l}}^{(1)} \bigg( 1 - \frac{\mu_{j, \setE_{l}}^{(1)}}{\sum \limits_{\ell \in \setL_{j}} \mu_{j, \setE_{\ell}}^{(1)} +\frac{\sigma^2}{B \rho}} \bigg) \bigg) \Big( B \sum \limits_{l \in \setL_{j}} \mu_{j, \setE_{l}}^{(1)} + \frac{\sigma^{2}}{\rho} \Big)}
\end{align}
\newcounter{countSINR}
\setcounter{countSINR}{\value{equation}}
\hrulefill
\end{figure*}

Our objective is to find the pilot reuse parameters $\beta$ and $\beta_{\mathrm{f}}$, as well as the number of scheduled users per cell $K$, that maximize the overall spectral efficiency within the cell. In the following, we obtain expressions for the spectral efficiency adopting MRC and P-ZFC combining. Without loss of generality, we will hereafter assume that the user coordinates $\z_{lk}$ are uniformly distributed within the cells.

\subsection{Achievable Spectral Efficiency}\label{subsec:achievable_spectral_efficiency}

We first give the following preliminary definitions. Let $\setI_{l}$ be the set of the $\beta_{\mathrm{f}} K$ users in cell $l \in \setL$ with the highest $d_{l}(\cdot)$; likewise, let $\setE_{l}$ be the set of the $(1 - \beta_{\mathrm{f}}) K$ users in cell $l \in \setL$ with the lowest $d_{l}(\cdot)$. Furthermore, we introduce the functions
\addtocounter{equation}{-7}
\begin{align}
\label{eq:mu_i} \mu_{j, \setI_{l}}^{(\gamma)} & \triangleq \Exp_{\z_{lk} \in \setI_{l}} \Big\{ \Big( \frac{d_{j}(\z_{lk})}{d_{l}(\z_{lk})} \Big)^{\gamma} \Big\} \\
\label{eq:mu_e} \mu_{j, \setE_{l}}^{(\gamma)} & \triangleq \Exp_{\z_{lk} \in \setE_{l}} \Big\{ \Big( \frac{d_{j}(\z_{lk})}{d_{l}(\z_{lk})} \Big)^{\gamma} \Big\}
\end{align}
for $\gamma=1,2$, where the ratio $d_{j}(\z_{lk}) / d_{l}(\z_{lk})$ expresses the relative strength of the interference received at BS $j$ from user $k$ in cell $l$.
Lastly, let $\setL_{j} \subseteq \setL$ denote the set of cells using the same subset of pilots as $\setE_{j}$.\footnote{Observe that $\setL_{j} = \setL$ occurs only when $\beta=1$.}

\begin{theorem}
The ergodic achievable spectral efficiency in cell $j$, when the users are uniformly distributed within the cell, is given by
\begin{align}
\label{eq:SE} & \mathrm{SE}_{j} \triangleq K \Big( 1 - \frac{B}{T} \Big) \\
\nonumber & \hspace{4mm} \times \Big( \beta_{\mathrm{f}} \log_{2} (1 + \mathsf{SINR}_{\setI_{j}}) + (1 - \beta_{\mathrm{f}}) \log_{2} (1 + \mathsf{SINR}_{\setE_{j}}) \Big)
\end{align}
with $B$ defined in \eqref{eq:FPR} and where the expressions of the signal-to-interference-plus-noise ratios (SINRs) are given in \eqref{eq:SINR_i_mrc}--\eqref{eq:SINR_e_mrc} for MRC and in \eqref{eq:SINR_i_zfc}--\eqref{eq:SINR_e_zfc} for P-ZFC, respectively, at the top of the page.
\end{theorem}

\begin{IEEEproof}
The expression for the ergodic achievable spectral efficiency in cell $j$ without FPR is given in \cite[Th.~1]{Bjo15}:
\setcounter{equation}{\value{countSINR}}
\begin{equation} \label{eq:SE_bjo}
\mathrm{SE}_{j} \triangleq \Big( 1 - \frac{B}{T} \Big) \sum_{k=1}^{K} \log_{2} (1 + \mathsf{SINR}_{jk}) 
\end{equation}
where $\mathsf{SINR}_{jk}$ is the SINR corresponding to user $k$ in cell $j$ derived in {\rm \cite[Th.~1,~2]{Bjo15}} for MRC and P-ZFC and
\begin{equation}
\mu_{j,l}^{(\gamma)} \triangleq \Exp_{\z_{lk}} \Big\{ \Big( \frac{d_{j}(\z_{lk})}{d_{l}(\z_{lk})} \Big)^{\gamma} \Big\}, \qquad \gamma = 1 ,2.
\end{equation}

We first note that, if the coordinates of all users within a cell have the same distribution, then $\mu_{j,l}^{(\gamma)}$ loses its (implicit) dependence on the user index $k$ and, as a consequence, we can rewrite the following term as
\begin{equation} \label{eq:sum_uniform}
\sum_{l \in \setL} \sum_{m=1}^{K} \mu_{j, l}^{(\gamma)} \v_{i_{jk}}^{\herm} \v_{i_{lm}} = \sum_{l \in \setL} \mu_{j, l}^{(\gamma)} \sum_{m=1}^{K} \v_{i_{jk}}^{\herm} \v_{i_{lm}} = B \sum_{l \in \setL_{j}} \mu_{j, l}^{(\gamma)}
\end{equation}
where the last equality derives from the fact that
\begin{equation}
\sum_{m=1}^{K} \v_{i_{jk}}^{\herm} \v_{i_{lm}} = \left\{
\begin{array}{ll}
B, & \quad \mathrm{if} \ l \in \mathcal{L}_j \\
0, & \quad \mathrm{otherwise.}
\end{array} \right.
\end{equation}
By extending \eqref{eq:sum_uniform} to the case of FPR, we can write
\begin{align}
\nonumber & \sum_{l \in \setL} \sum_{m=1}^{K} \mu_{j, l}^{(\gamma)} \v_{i_{j k}}^{\herm} \v_{i_{jk}} = \\
& = \sum_{l \in \setL} \sum_{m \in \setI_{l}} \mu_{j, \setI_{l}}^{(\gamma)}\v_{i_{jk}}^{\herm} \v_{i_{lm}}+ \sum_{l \in \setL} \sum_{m \in \setE_{l}} \mu_{j, \setE_{l}}^{(\gamma)} \v_{i_{jk}}^{\herm} \v_{i_{lm}} \\
& = \sum_{l \in \setL} \mu_{j, \setI_{l}}^{(\gamma)} \sum_{m \in \setI_{l}} \v_{i_{jk}}^{\herm} \v_{i_{lm}} + \sum_{l \in \setL}\mu_{j, \setE_{l}}^{(\gamma)}\sum_{m \in \setE_{l}} \v_{i_{jk}}^{\herm} \v_{i_{lm}} \\
\label{eq:sum_frac} & = B \Big( I_{\setI_{j}}(k) \sum_{l \in \setL} \mu_{j, \setI_{l}}^{(\gamma)} + \big( 1 - I_{\setI_{j}}(k) \big) \sum_{l \in \setL_{j}} \mu_{j, \setE_{l}}^{(\gamma)} \Big)
\end{align}
where we have defined
\begin{equation}
I_{\setI_{j}}(k) \triangleq \left\{
\begin{array}{ll}
1, & \quad \mathrm{if} \ k \in \setI_{j} \\
0, & \quad \mathrm{if} \ k \in \setE_{j}
\end{array} \right.
\end{equation}
and we obtain the SINRs in \eqref{eq:SINR_i_mrc}--\eqref{eq:SINR_e_zfc}. Finally, exploiting the fact that $|\setI_{j}| = \beta_{\mathrm{f}} K$, we arrive at the expression of the spectral efficiency in \eqref{eq:SE}.
\end{IEEEproof}

\begin{figure}[t!]
\centering
\includegraphics[scale=1]{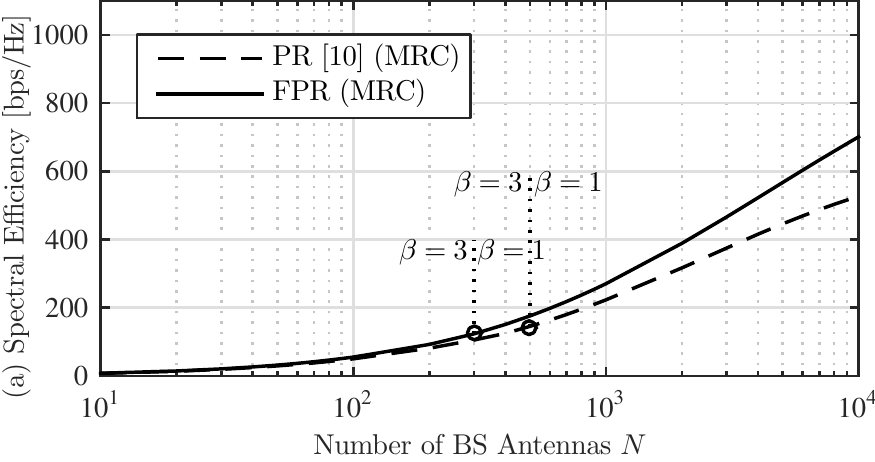} \\
\includegraphics[scale=1]{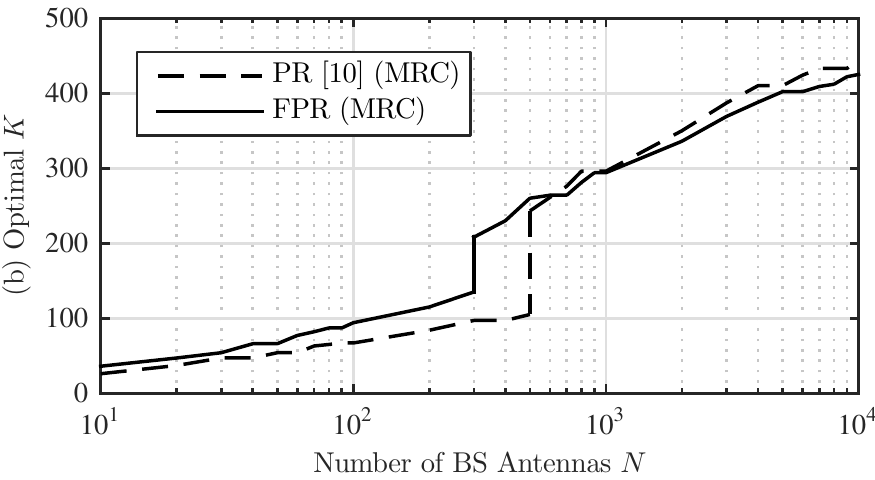}
\caption{Comparison between pilot reuse (PR) \cite{Bjo14} and fractional pilot reuse (FPR) with MRC: (a) spectral efficiency of the cell as a function of $N$; (b) Optimal number of scheduled users as a function of $N$.} \label{fig:FPR_mrc}
\end{figure}

\vspace{2mm}

\noindent It is straightforward to conclude that, when $\beta_\mathrm{f} = 0$, we have $\setI_{l} = \emptyset$ and the spectral efficiency in \eqref{eq:SE} collapses to the expression given in \eqref{eq:SE_bjo}.

\begin{remark} {\rm
The expressions above can be easily extended to the case where the $K$ mobile users are not uniformly distributed over the cell at the cost of less tractability. More specifically, when the average relative strength of a user depends on its index, then the terms $\mu_{j, \setI_{l}}^{(\gamma)}$ and $\mu_{j, \setE_{l}}^{(\gamma)}$ cannot be removed from the inner summation. Nonetheless, the resulting formulas can be evaluated by Monte-Carlo simulations if required.}
\end{remark}

\begin{figure}[t!]
\centering
\includegraphics[scale=1]{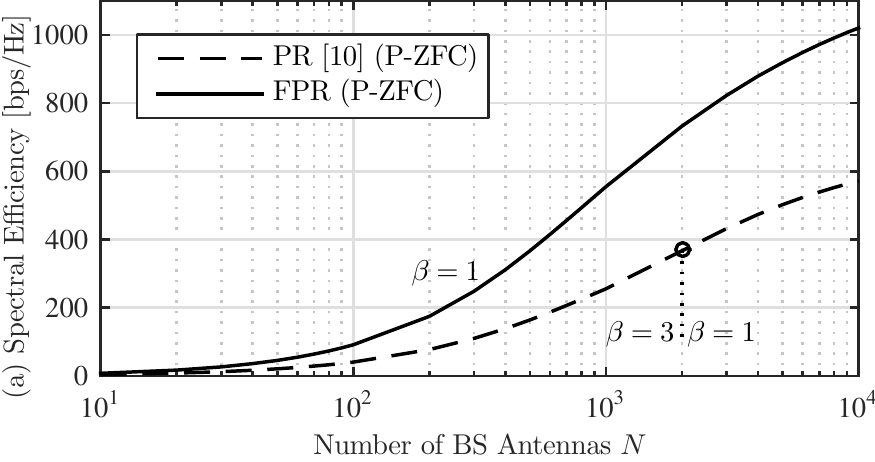} \\
\includegraphics[scale=1]{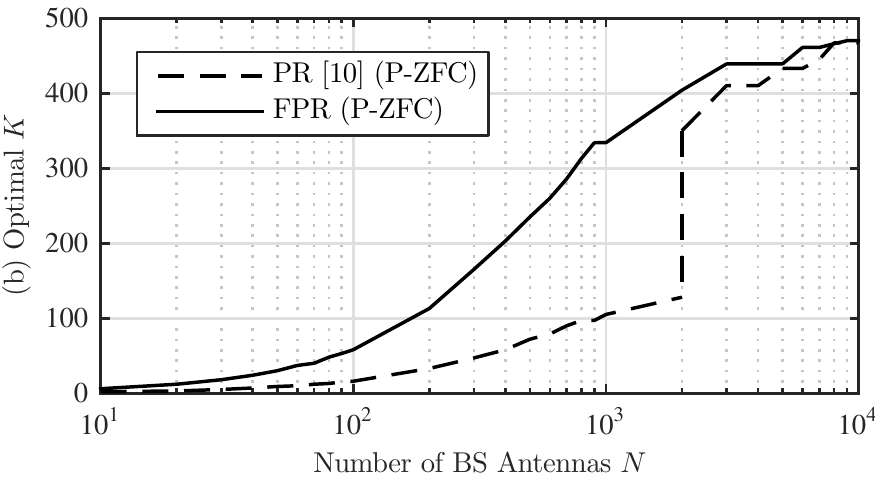}
\caption{Comparison between pilot reuse (PR) \cite{Bjo14} and fractional pilot reuse (FPR) with P-ZFC: (a) spectral efficiency of the cell as a function of $N$; (b) Optimal number of scheduled users as a function of $N$.} \label{fig:FPR_zfc}
\end{figure}

Now, let us study the performance of the proposed scheme based on FPR when $N \to \infty$.

\begin{corollary} \label{cor:asympt}
The spectral efficiency in \eqref{eq:SE} in the large-antenna regime ($N \to \infty$), when the users are uniformly distributed within the cell, becomes
\begin{align} \label{eq:SE_lim}
\nonumber \lim_{N \to \infty} \mathrm{SE}_{j} = \ & K \Big( 1 - \frac{B}{T} \Big) \Bigg( \beta_\mathrm{f} \log_2 \Bigg( 1 + \frac{1}{\sum \limits_{l \in \setL  \backslash \{j\}} \mu_{j, \setI_{l}}^{(2)}} \Bigg) \\
& + (1 - \beta_\mathrm{f}) \log_2 \Bigg( 1 + \frac{1}{\sum \limits_{l \in \setL_{j} \backslash \{j\}} \mu_{j, \setE_{l}}^{(2)}}\Bigg) \Bigg)
\end{align}
with $B$ defined in \eqref{eq:FPR}.
\end{corollary}

\noindent Evidently, the asymptotic spectral efficiency derived in Corollary~\ref{cor:asympt} does not depend on the type of combining employed at the BS. Again, we observe that, when $\beta_\mathrm{f} = 0$, the same expression of the asymptotic spectral efficiency given in \cite[Cor.~3]{Bjo15} is obtained.

Unfortunately, as in \cite{Bjo15}, the expressions obtained in \eqref{eq:SE} and in \eqref{eq:SE_lim} do not allow for closed-form extraction of the optimal $\{ \beta, \beta_{\mathrm{f}}, K \}$; moreover, a further complication of our scheme is that the terms $\mu_{j, \setI_{l}}^{(\gamma)}$ and $\mu_{j, \setE_{l}}^{(\gamma)}$ intrinsically depend on $\beta_{\mathrm{f}}$. Therefore, in the next section, we optimize the system parameters using numerical evaluation for a particular scenario.

\section{Numerical results} \label{sec:numerical_results}

In this section, we assess the performance of the proposed scheme based on FPR through numerical evaluation. In doing so, we adopt an equivalent setup to \cite{Bjo14} and compare our results against those obtained without using FPR.

Let us consider a symmetric, infinitely large network of hexagonal cells with radius $r > 0$. Furthermore, let us assume a standard pathloss model in which the variance of the channel attenuation is given by $d_{j}(\z) = \| \z - \b_{j} \|_{2}^{-\kappa}$, where $\kappa \geq 2$ is the pathloss exponent and $\b_{j} \in \Real^{2}$ specifies the coordinates of BS $j$: in this context, we have (cf. \eqref{eq:mu_i}--\eqref{eq:mu_e})
\begin{equation}
\frac{d_{j}(\z_{lk})}{d_{l}(\z_{lk})} = \Big( \frac{\| \z_{lk} - \b_{l} \|_{2}}{\| \z_{lk} - \b_{j} \|_{2}} \Big)^{\kappa}.
\end{equation}

\begin{figure}[t!]
\centering
\includegraphics[scale=1]{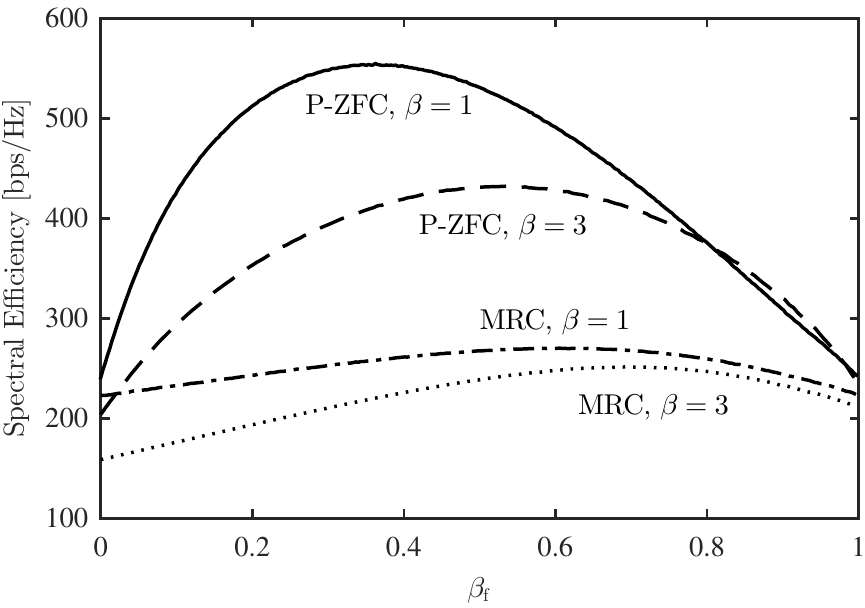}
\caption{Spectral efficiency with $N=1000$ antennas as a function of $\beta_{\mathrm{f}}$.} \label{fig:beta_f1}
\end{figure}

\begin{table}[t!]
\centering
\renewcommand*{\arraystretch}{1.3}
\begin{tabular}{llrrr}
\vspace{1mm} \\
\hline
		 & & MRC        & & P-ZFC \\
\hline
$N=10^1$ & & $5.6\,\%$  & & $93.1\,\%$ \\
$N=10^2$ & & $10.4\,\%$ & & $125.8\,\%$ \\
$N=10^3$ & & $21.2\,\%$ & & $117.4\,\%$ \\
$N=10^4$ & & $28.6\,\%$ & & $81.3\,\%$ \\
\hline
\end{tabular}
\vspace{3mm}
\caption{Gains of FPR with respect to \cite{Bjo14}.} \label{table:gains}
\vspace{-7mm}
\end{table}

For our numerical simulations, the same parameters as in \cite{Bjo14} are used: we take into account the interference produced by $3$ tiers of surrounding cells, a coherence block of $T=1000$ channel uses, a pathloss exponent $\kappa=3.5$, an average SNR between any user and any antenna at the corresponding BS of $\rho/\sigma^{2} = 10$ dB, and a number of BS antennas in the range $N \in [10, 10^4]$. Lastly, the terms $\mu_{j, \setI_{l}}^{(\gamma)}$ and $\mu_{j, \setE_{l}}^{(\gamma)}$ are computed by Monte-Carlo simulations as follows. For each value of $K$, we generate $10^{6}$ instances of $K$ uniformly distributed user coordinates in every cell $l$ with a minimum distance of $0.14 r$ from the corresponding BS; then, for each value of $\beta_{\mathrm{f}}$, i) the $\beta_{\mathrm{f}} K$ coordinates that are closest to the BS are assigned to $\setI_{l}$ and $\mu_{j, \setI_{l}}^{(\gamma)}$ is computed as in \eqref{eq:mu_i}, and ii) the other coordinates are assigned to $\setE_{l}$ and $\mu_{j, \setE_{l}}^{(\gamma)}$ is computed as in \eqref{eq:mu_e}.

Remarkably, using FPR always improves the spectral efficiency with respect to \cite{Bjo14}, as summarized in Table~\ref{table:gains}. Figure~\ref{fig:FPR_mrc}(a) and Figure~\ref{fig:FPR_mrc}(b) show the spectral efficiency and optimal number of scheduled users, respectively, when using MRC combining; despite obtaining quite similar optimal values of $K$ with respect to the baseline \cite{Bjo14}, we observe that adopting FPR allows to anticipate the switching point from $\beta=3$ to $\beta=1$, i.e., the channels become quasi-orthogonal at lower values of $N$. Likewise, Figure~\ref{fig:FPR_zfc} illustrates the same results for the case of P-ZFC, which produces significantly improved spectral efficiencies with respect to MRC. Note that, with P-ZFC, choosing $\beta=1$ yields a higher spectral efficiency with respect to $\beta=3$ even for low values of $N$, which always allows to schedule more users than without FPR (see Figure~\ref{fig:FPR_zfc}(b)). 

Let us more generally compare MRC and P-ZFC: in the former, the impact of our approach is more significant as the number of antennas at the BS increases whereas, for the latter, our method exceptionally doubles the spectral efficiency in the whole range of $N$ under exam. In addition, by comparing Figure~\ref{fig:FPR_mrc}(a) and Figure~\ref{fig:FPR_zfc}(a), it emerges that FPR with MRC always outperforms the baseline \cite{Bjo14} with P-ZFC.

\begin{figure}[t!]
\centering
\includegraphics[scale=1]{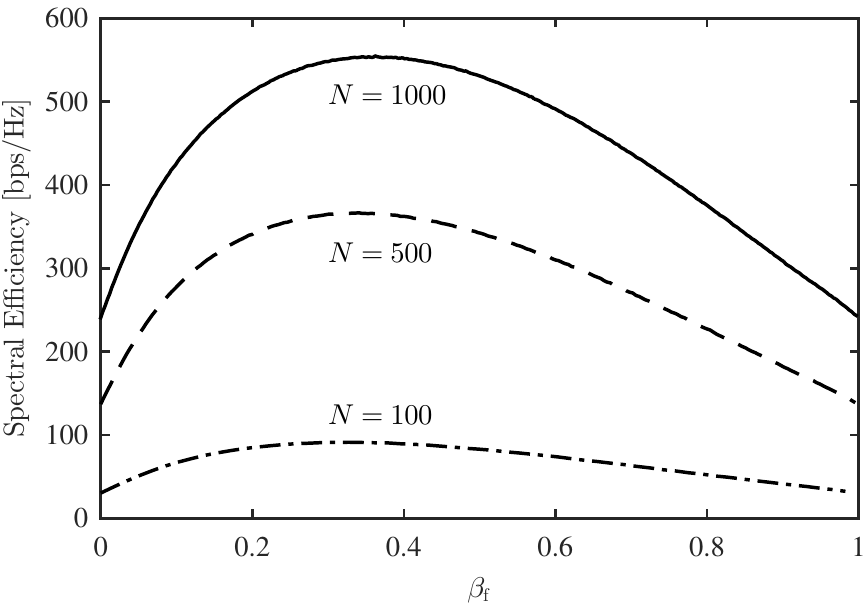}
\caption{Spectral efficiency using P-ZFC and $\beta=1$ as a function of $\beta_{\mathrm{f}}$ for different values of $N$.} \label{fig:beta_f2}
\end{figure}

As expected, the spectral efficiency in \eqref{eq:SE} is concave in $\beta_{\mathrm{f}}$ for any fixed $K$. The spectral efficiency as a function of $\beta_{\mathrm{f}}$ is depicted in Figure~\ref{fig:beta_f1} with $N=10^3$ BS antennas and using the optimal value of $K$ for each setup: in particular, we observe that the optimal $\beta_{\mathrm{f}}$ changes with both the integer reuse factor $\beta$ and the combining technique. Lastly, the spectral efficiency obtained with P-ZFC $\beta=1$, i.e., the best combination of integer reuse factor and combining technique, is illustrated as a function of $\beta_{\mathrm{f}}$ in Figure~\ref{fig:beta_f2} for some values of $N$.

\section{Conclusions} \label{sec:conclusions}

In this paper, we propose a novel pilot allocation scheme for multi-cell massive MIMO systems, where users close to their respective base stations get to reuse the same set of pilot sequences across the whole system. We derive expressions for an achievable rate in this scenario and obtain both the pilot reuse parameters and the number of scheduled users per cell that maximize it. Results show a significant gain in terms of spectral efficiency with respect to the existing pilot reuse baseline.

\section*{Acknowledgements}

The authors would like to thank Emil Bj\"{o}rnson and Marios Kountouris for their valuable comments and suggestions.

\clearpage

\addcontentsline{toc}{chapter}{References}
\bibliographystyle{IEEEtran}
\bibliography{IEEEabrv,ref_Huawei}

\end{document}